\documentclass[twocolumn,secnumarabic,amssymb, nobibnotes, aps, prl, superscriptaddress]{revtex4-2}
\usepackage{booktabs}
\usepackage{enumitem}
\usepackage{graphicx}% Include figure files
\usepackage{dcolumn}% Align table columns on decimal point
\usepackage{bm}% bold math
\usepackage{amsmath}
\usepackage{ragged2e}
\usepackage{enumerate}
\usepackage{setspace}
\usepackage{dsfont}
\usepackage{physics}
\usepackage{subfigure}
\usepackage{multirow}
\usepackage{indentfirst} %段落缩进值宏包
\usepackage {mathrsfs}%花体字
\usepackage{color}%字体颜色
\usepackage[margin=2.3cm]{geometry}
\usepackage{amssymb,amsfonts,amsthm,bm,mathtools}
\usepackage{makecell}   %表格
\usepackage{multirow}	%表格

\usepackage{xcolor}  %颜色

\usepackage{threeparttable}  %表格注释
\usepackage[colorlinks=true,linkcolor=blue,citecolor=blue]{hyperref}

\theoremstyle{definition}

\begin{document}

\title{An optical-frequency-comb-enabled unified infrastructure for quantum networks}%
\author{Yiting Lin}%
\affiliation{Beijing Key Laboratory of Quantum Sensing and Precision Measurement, and Center for Quantum Information Technology, and Institute of Quantum Electronics, Peking University, Beijing 100871, China}

\author{Zishuo Ren}%
\affiliation{Beijing Key Laboratory of Quantum Sensing and Precision Measurement, and Center for Quantum Information Technology, and Institute of Quantum Electronics, Peking University, Beijing 100871, China}

\author{Yunkun Zhao}%
\affiliation{State Key Laboratory of Environment Characteristics and Effects for Near-space, Center for Photonic Quantum Precision Measurement, Advanced Research Institute of Multidisciplinary Science, Beijing Institute of Technology, Beijing 100081, China}
\affiliation{School of Optics and Photonics, Beijing Institute of Technology, Beijing 100081, China}

\author{Jiayu Ma}%
\affiliation{State Key Laboratory of Information Photonics and Optical Communications, Beijing University of Posts and Telecommunications, Beijing 100876, China}

\author{Zimeng Li}%
\affiliation{Beijing Key Laboratory of Quantum Sensing and Precision Measurement, and Center for Quantum Information Technology, and Institute of Quantum Electronics, Peking University, Beijing 100871, China}

\author{Yang Sun}%
\affiliation{Technology and Engineering Center for Space Utilization, Chinese Academy of Sciences, 100094, China}

\author{Yufei Zhang}%
\affiliation{Beijing Key Laboratory of Quantum Sensing and Precision Measurement, and Center for Quantum Information Technology, and Institute of Quantum Electronics, Peking University, Beijing 100871, China}

\author{Yanhao Sun}%
\affiliation{State Key Laboratory of Information Photonics and Optical Communications, Beijing University of Posts and Telecommunications, Beijing 100876, China}

\author{Lei Yang}%
\affiliation{Technology and Engineering Center for Space Utilization, Chinese Academy of Sciences, 100094, China}

\author{Xiangyu Wang}%
\affiliation{State Key Laboratory of Information Photonics and Optical Communications, Beijing University of Posts and Telecommunications, Beijing 100876, China}

\author{Kemi Xu}%
%\email[E-mail: ]{xukemi@bit.edu.cn}
\affiliation{State Key Laboratory of Environment Characteristics and Effects for Near-space, Center for Photonic Quantum Precision Measurement, Advanced Research Institute of Multidisciplinary Science, Beijing Institute of Technology, Beijing 100081, China}
\affiliation{School of Optics and Photonics, Beijing Institute of Technology, Beijing 100081, China}

\author{Ziyang Chen}%
\email[E-mail: ]{chenziyang@pku.edu.cn}
\affiliation{Beijing Key Laboratory of Quantum Sensing and Precision Measurement, and Center for Quantum Information Technology, and Institute of Quantum Electronics, Peking University, Beijing 100871, China}
\affiliation{State Key Laboratory of Environment Characteristics and Effects for Near-space, Center for Photonic Quantum Precision Measurement, Advanced Research Institute of Multidisciplinary Science, Beijing Institute of Technology, Beijing 100081, China}

\author{Hong Guo}%
\email[E-mail: ]{hongguo@pku.edu.cn}
\affiliation{Beijing Key Laboratory of Quantum Sensing and Precision Measurement, and Center for Quantum Information Technology, and Institute of Quantum Electronics, Peking University, Beijing 100871, China}
\affiliation{State Key Laboratory of Environment Characteristics and Effects for Near-space, Center for Photonic Quantum Precision Measurement, Advanced Research Institute of Multidisciplinary Science, Beijing Institute of Technology, Beijing 100081, China}

\begin{abstract} 

Quantum networks require precision timing and classical communication to coordinate distributed quantum operations, yet these functions are typically supported by separate infrastructures, increasing system complexity and limiting scalability. Here we demonstrate an optical-frequency-comb-enabled infrastructure that unifies precision timing, classical communication and quantum communication over a shared fibre link. By encoding classical data onto the intensity of the comb while retaining its pulse timing, we simultaneously transmit data at 100~Mbit~s$^{-1}$ and achieve a timing stability of 0.85~fs at 1~s over 100~km. We develop a spectral-engineering strategy to suppress the coupling between data modulation and timing extraction. We further establish a power-dependent model of the excess noise induced in a co-propagating continuous-variable quantum key distribution channel, identifying optical power, wavelength allocation and propagation direction as the principal control parameters. After optimization, the system introduces only 0.0037~shot noise unit of additional excess noise and limits the secret-key-rate penalty to 6.7\%. Furthermore, quantum-secured timestamp transmission additionally enables absolute time synchronization with a stability of 0.26~fs at 1~s. These results establish optical frequency combs as a common physical layer for integrated quantum-network services.

\end{abstract}

\maketitle
%\tableofcontents  %目录

%\section{INTRODUCTION}\label{sec1}
   \begin{figure*}[!t]
	\centering
	\includegraphics[width= \linewidth]{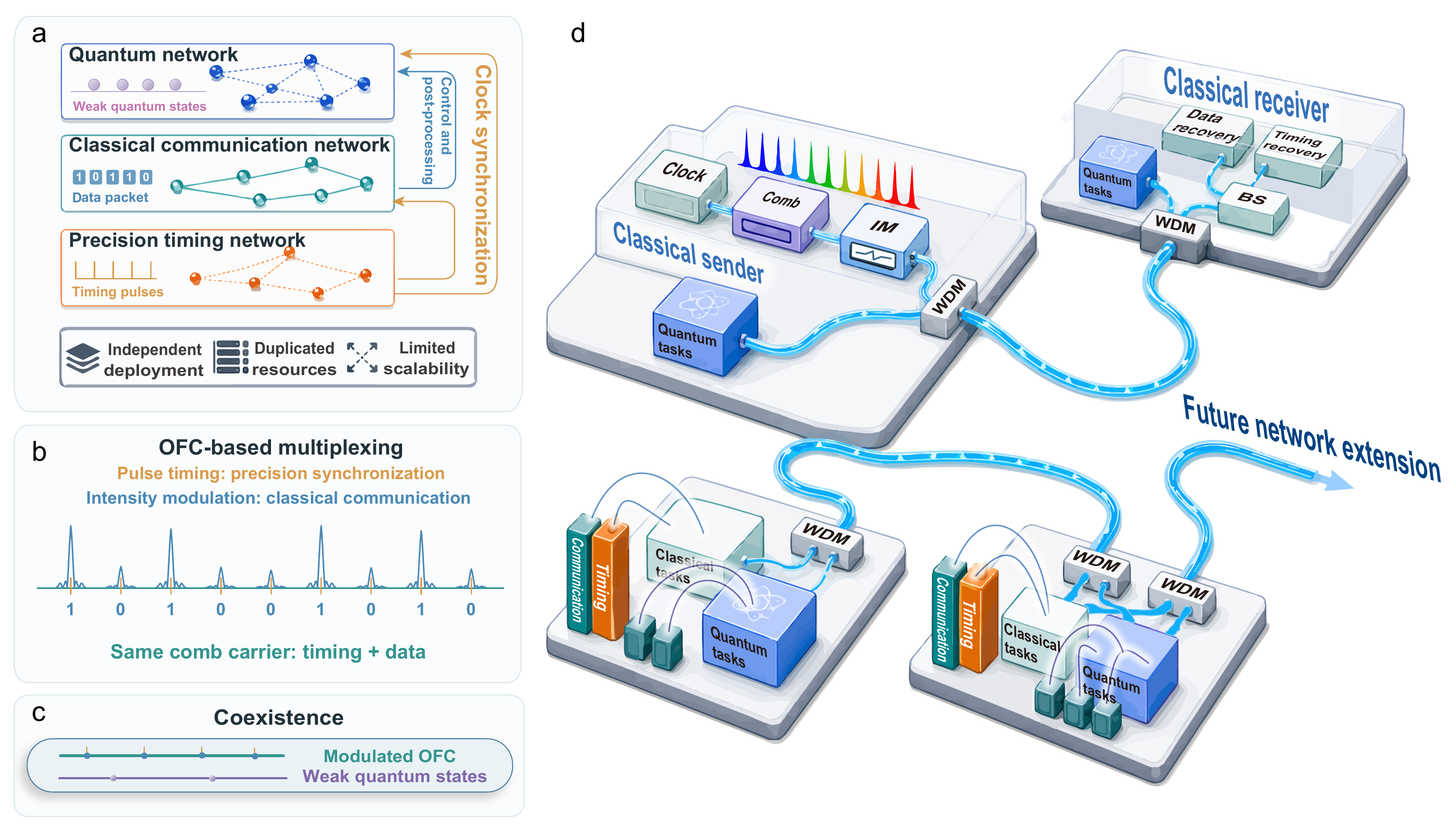}
	\caption{Concept and architecture of an optical-frequency-comb-enabled unified infrastructure for quantum networks. a, Conventional quantum networks rely on separate classical communication and precision timing networks for functions such as control, post-processing and synchronization. b,  
    The optical frequency comb (OFC) carries precision timing through its pulse timing and classical communication through intensity modulation. c, The OFC and weak quantum states are multiplexed and co-transmitted through wavelength-division multiplexing. d, The unified quantum-network infrastructure proposed in this work. WDM, wavelength-division multiplexing; IM, intensity modulator; BS, beam splitter.}\label{Fig1}
    \end{figure*}

% 量子网络是未来重要基础设施，其稳定运行，需要经典网络提供时钟同步和认证，来处理复杂的量子任务，例如分布式量子计算、量子传感和量子通信。然而，目前量子网络和经典网络相对独立且分离，这会导致未来量子网络需要依赖复杂但需要精确协同的三张分立网络，结构复杂，且资源利用率低、可扩展性差。而直接将三张网络融合，会因为物理层不兼容、各网络对信号的功率和噪声要求存在矛盾而难以实现。这自然引出问题：有没有一种方法，可以将三张分立网络协同起来？
Quantum networks are expected to enable fundamentally new capabilities beyond classical networks, including secure communication, distributed quantum computing and enhanced quantum sensing 
~\cite{wang2025laser,malia2022distributed,guo2020distributed,wehner2018quantum,marra2018ultrastable,komar2014quantum,kimble2008quantum,cirac1999distributed}. 
Recent advances in quantum communication, quantum processors and quantum interfaces have demonstrated key building blocks for such networks~\cite{huang2026fusion,main2025distributed,pittaluga2025deployed,knaut2024entanglement,hermans2022teleportation,liu2021distributed,pompili2021multinode,chen2021spaceground,daiss2021quantumgate}. However, realizing practical and scalable quantum networks requires more than quantum information processing technologies themselves. Similar to today’s classical Internet, future quantum networks will rely on a complex infrastructure consisting of quantum channels, classical communication networks and high-precision timing systems to coordinate distributed quantum operations~\cite{delledonne2025operating,burenkov2023synchronization}.

At present, these infrastructures are developed largely independently. %%
Quantum communication systems require classical channels for control and post-processing, while distributed quantum operations rely on precise timing references provided by dedicated synchronization networks~\cite{mckenzie2024clock,knaut2024entanglement,alshowkan2022advanced,cuomo2020ecosystem,wehner2018quantum}, as shown in Fig.~\ref{Fig1}a. The coexistence of multiple independent networks introduces substantial challenges in scalability, resource efficiency and system complexity. Moreover, direct integration of these networks at the signal level remains difficult because quantum signals, classical optical signals and precision timing signals have distinct requirements in terms of optical power and noise tolerance~\cite{iqbal2025sdn,burenkov2023synchronization}. Therefore, a key challenge for future quantum networks is to develop a unified infrastructure that can support timing, communication and quantum applications within a compatible physical framework.

% 光梳是支持高精度时频传递的载体，其能实现的超高精度时频传递是其他方法难以实现的。其多自由度的属性允许我们利用其强度维度携带通信数据，从而实现时频信息和通信数据在统一载体上同传。这种方法可以减少量子网络同时对两个网络的依赖，而只需与时频-通信网融合，这可以给量子任务提供一个良好的经典网络支撑。
Optical frequency combs (OFCs) provide a promising route towards such an integrated infrastructure~\cite{zheng2026large,chen2025time,chang2022integrated,diddams2020combs}. Owing to their intrinsically high temporal resolution and multiple degrees of freedom
, optical frequency combs have enabled breakthroughs in precision timing and ranging~\cite{chen2024dualcomb,yu2023time,shen2022freespace,caldwell2022time,wang2018subshot,deschenes2016synchronization,jung2014remote,giorgetta2013optical}, frequency metrology~\cite{yang2025frequencymetrology,yu2024microwave,schioppo2022comparing,jung2014frequency} and high-capacity optical communication~\cite{jorgensen2022petabit,marinpalomo2017communications}. 
Here, we explore a new role of optical frequency combs as a common physical platform for quantum network infrastructure, rather than merely as a precision measurement tool. By encoding classical communication information onto the intensity domain of optical comb signals while preserving their timing information, timing and communication functions can be fused into a shared optical carrier, as illustrated in Fig.~\ref{Fig1}b. Such an approach can reduce the dependence on parallel supporting networks and provide a compatible foundation for quantum tasks.

% 然而，将通信与时频融合在同一载体上面临最基本的问题：通信会让光梳承载时频信息的能量中的一部分转移为通信谱，通信谱和时频谱的混叠会让时频谱的识别存在困难，从而引起时频传递稳定度的下降。此外，这种高峰值功率的宽的复杂谱经典光与量子光共传缺少研究，但其总功率、峰值功率都高的特点必然会对量子信道产生影响，这是需要被定量描述的。
However, integrating communication functions with precision timing introduces fundamental challenges. Optical modulation used for data transmission inevitably generates spectral sidebands, which may overlap with the timing information carried by optical combs and degrade synchronization performance. Furthermore, as illustrated in Fig.~\ref{Fig1}c, introducing high-peak-power classical optical signals into the same fibre infrastructure as weak quantum states may induce excess noise through phase distortion and photon contamination, which arise from Kerr nonlinearities~\cite{du2020fwm,kumar2015coexistence}, Raman scattering~\cite{ferreira2014raman} and optical power leakage due to imperfect channel isolation~\cite{eriksson2019wdm,qi2010feasibility}. Therefore, achieving a practical fused infrastructure requires both suppressing timing–communication coupling and ensuring compatibility with quantum channels. 

% 为了解决这些问题，我们提出了一种光梳赋能量子网络的物理基础设施，通过选择合适的通信策略，即选择合适的码型和码率、调制深度，来减弱通信谱和调制谱的混叠，并在实验上对其可行性做了验证；对具有复杂谱的经典光对量子信道的影响做定量分析，揭示了其对量子信道的影响的变化规律，并在100 km光纤链路上验证了共传对CV-QKD的影响可限制在xx % 以内。我们还引用这种架构，做了信息论安全的绝对时间同步，验证其基本对时频传递的稳定度没有影响。
Here we demonstrate an optical-frequency-comb-enabled infrastructure that unifies precision timing, classical communication and quantum communication over a shared fibre link, as illustrated in Fig.~\ref{Fig1}d. By encoding classical data onto the intensity of the comb while retaining its pulse timing, we simultaneously transmit 100~Mbit~s$^{-1}$ classical data and achieve a timing stability of 0.85~fs at 1~s over 100~km. We further develop a spectral-engineering strategy that controls the modulation rate, symbol distribution and modulation depth to minimize the coupling between data modulation and timing extraction. To integrate quantum communication, we establish a power-dependent model of the excess noise induced by the broadband, high-peak-power comb and identify optical power, wavelength allocation and propagation direction as the principal control parameters. With these optimizations, the infrastructure coexists with continuous-variable quantum key distribution (CV-QKD) over the same 100-km link, introducing only 0.0037~shot noise unit (SNU) of additional excess noise and limiting the secret-key-rate penalty to 6.7\%. Finally, we achieve quantum-secured timestamp transmission enabling absolute time synchronization with a stability of 0.26~fs at 1~s. These results establish optical frequency combs as a common physical layer for the timing, communication and security functions required by scalable quantum networks.

    \begin{figure*}%[t]
	\centering
	\includegraphics[width= \linewidth]{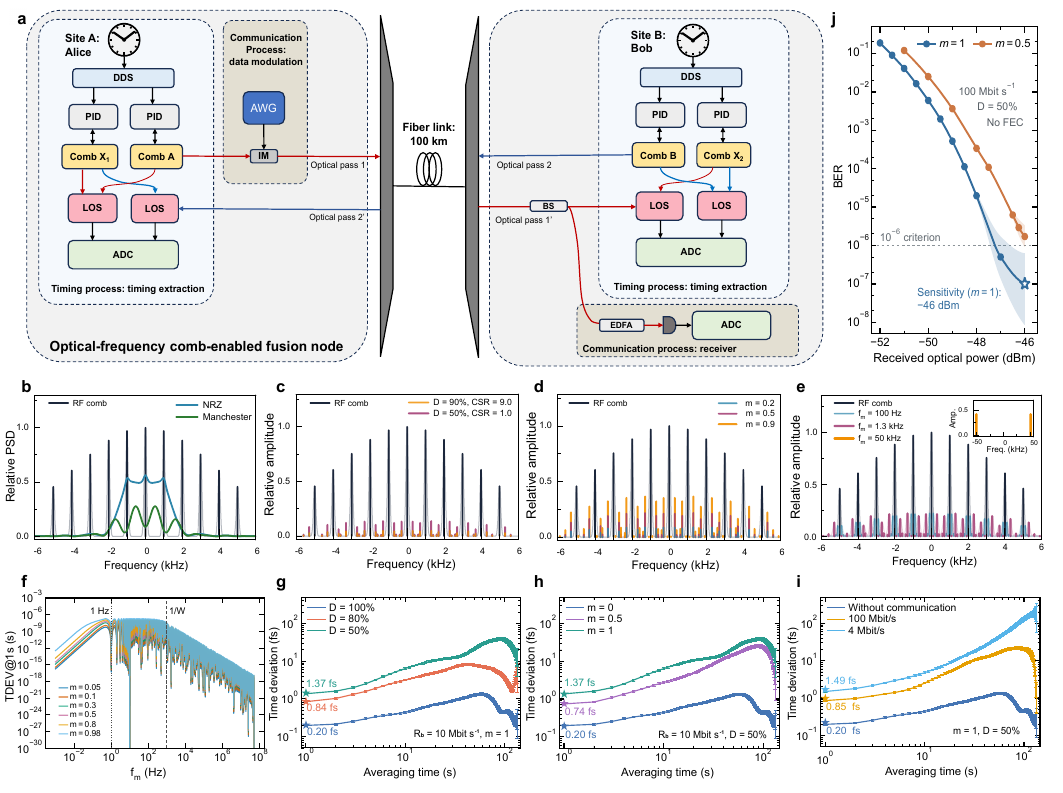}
	\caption{Fusion of precision timing and classical communication enabled by optical frequency combs. a, Experimental architecture. 
    The repetition frequency of the optical frequency comb is locked to an atomic clock to transfer accurate time--frequency information, while classical communication data are encoded in the intensity domain through intensity modulation. 
    DDS, direct digital synthesizer; PID, proportional–integral–derivative controller; LOS, linear optical sampling; AWG, arbitrary waveform generator; ADC, analogue-to-digital converter; EDFA, erbium-doped fibre amplifier.
    b-e, Simulated spectra showing how different modulation strategies reshape the position and power of communication--induced spectral components relative to the timing-sensitive region. The black lines denote the radio-frequency (RF) comb generated by dual-comb beating for time--frequency recovery, while the coloured ones are communication--induced components. The vertical axes give  the relative power spectral density (PSD) in b and the relative spectral amplitude in c--e.
    b, 
    For NRZ and Manchester coding. 
    c, 
    For unmodulated-symbol fraction $D=90\%$ (CSR$=9.0$) and $D=50\%$ (CSR$=1.0$). 
    d, 
    For modulation depths $m=0.2$, $0.5$ and $0.9$.
    e, 
    For $f_{\mathrm{m}}=100~\mathrm{Hz}$, $1.3~\mathrm{kHz}$ and $50~\mathrm{kHz}$, where $f_{\mathrm{m}}$ denotes the modulation frequency and increases with the bit rate $R_b$. 
    f, Theoretical analysis of the effects of modulation rate and modulation depth on timing stability. 
    g-i, Experimental characterization of the effect of  $D$, $m$ and $R_b$ on timing stability. 
    From bottom to top, the curves correspond to  $D=100\%$, $80\%$ and $50\%$ ($R_b = 10~\mathrm{Mbit~s}^{-1}, m = 1$) in g;  $m=0$, $0.5$ and $1$ ($R_b = 10~\mathrm{Mbit~s}^{-1}, D = 50\%$) in h; and the unmodulated case, 100 and 4~Mbit~s$^{-1}$ ($m = 1, D = 50\%$)   in i. 
    Error bars in g–i indicate one standard error of the TDEV estimates. 
    j, Bit error rate (BER) versus received optical power at 100 Mbit s\(^{-1}\), $D=50\%$ for \(m=1\) and \(m=0.5\). 
    Shaded regions indicate 95$\%$ confidence intervals.  The star marks the \(-46\) dBm sensitivity at BER \(=10^{-6}\) without forward error correction (FEC). 
    }\label{Fig2}
    \end{figure*}

\section*{Results}
\subsection*{Optical-frequency-comb-enabled fusion of timing and communication infrastructures}
Optical frequency combs naturally provide multiple degrees of freedom for information encoding. The precisely controlled pulse timing enables high-resolution time-frequency transfer, while the optical intensity can be modulated to encode classical communication data. By combining these two degrees of freedom, timing information and communication data can be simultaneously transmitted through the same optical carrier without requiring an additional communication channel (see Methods for the principle of timing synchronization and the timing--communication fusion).

A practical quantum network requires not only quantum information processing nodes, but also supporting infrastructures that provide accurate timing references and classical information exchange. Conventionally, precision timing and classical communication are implemented through independent optical networks, resulting in additional hardware complexity and limited scalability. Here, we demonstrate the fusion of these two essential infrastructures by using an optical frequency comb as a shared physical carrier, as shown in Fig.~\ref{Fig2}a (see Methods for detailed system architecture).

The key challenge of such a fused architecture is the interaction between communication modulation and timing extraction. Intensity modulation introduces spectral sidebands around comb modes, which may overlap with the spectral components used for precision timing measurement and appear as additional timing noise because timing is recovered from the modulated dual-comb interferogram. 

To overcome this limitation, we develop a spectral-engineering strategy that exploits the relationship between modulation  and timing characteristics. By optimizing the coding formats, symbol distribution, modulation depth and  modulation rate, the communication--induced dominant 
spectral components can be separated from the timing information and their contribution to timing instability can be minimized. 
As shown in Fig.~\ref{Fig2}b, different coding formats reshape the modulation spectrum. For example,  compared with non-return-to-zero (NRZ), Manchester coding suppresses low-frequency components and redistributes the spectral power towards higher frequencies, reducing the spectral overlap at the expense of communication efficiency. 
Meanwhile, the energy of communication--induced components can be further controlled through the fraction of unmodulated symbols $D$ and the modulation depth $m$. A higher $D$, as in the yellow lines in Fig.~\ref{Fig2}c, increases the carrier-to-sideband power ratio (CSR), and a smaller $m$, as in the blue lines in Fig.~\ref{Fig2}d, suppresses the sideband power. These controls therefore provide a trade-off between suppressing timing--communication coupling and maintaining communication performance, such as 
communication efficiency or sufficiently low bit-error rate (BER).
Furthermore, as Fig.~\ref{Fig2}e shows, increasing the modulation frequency $f_{\mathrm{m}}$ shifts the sidebands away from 
the spectral region carrying the time--frequency information, as illustrated by the 50~kHz case,  thereby reducing interference. 

Theoretical calculations (as shown in Fig.~\ref{Fig2}f; see Supplementary Note~1 for the derivation) quantify the dependence of timing stability on \(f_{\mathrm{m}}\) and \(m\), with stability characterized by time deviation (TDEV) at 1~s. 
As \(f_{\mathrm{m}}\) increases, the TDEV first rises in the very-low-frequency region and then exhibits an oscillatory, weakly varying regime owing to the finite temporal window $W$ used in the practical timing-extraction method. Above the characteristic scale set by \(1/W\), however, the TDEV exhibits an overall decreasing trend with increasing \(f_{\mathrm{m}}\). The former two regimes correspond to modulation rates far below those relevant to practical communication. Hence, within the practical communication range, the communication--induced degradation of timing stability diminishes as the modulation rate increases.  Meanwhile, reducing \(m\) further shifts the TDEV downward, resulting in a smaller degradation of timing stability.
Furthermore,  as illustrated in Fig.~\ref{Fig2}g--i, experiments reinforce that these controls enable effective decoupling between communication data and timing information for principle validation. 

Using this spectral-engineering strategy, we experimentally realize a fused time--frequency communication link based on an optical frequency comb.  We use NRZ coding and set $D=50\%$ to maintain communication efficiency, and set $m=1$ to ensure a low BER.  The system simultaneously transfers precision timing signals and classical communication data over a 100-km fibre link, achieving a timing stability of 0.85 fs at 1~s while supporting a communication rate of 100 Mbit~$\mathrm{s}^{-1}$, as shown in Fig.~\ref{Fig2}i. 
Meanwhile, it meets the \(10^{-6}\) BER criterion at a received optical power of -46 dBm at 100 Mbit s\(^{-1}\) without forward-error correction, as shown in Fig.~\ref{Fig2}j. 
It can also be seen that OFCs, as high signal-to-noise ratio light sources, provide a promising approach for low BER communication. 
These results demonstrate simultaneous high-speed communication and sub-femtosecond timing stability over a shared optical carrier. Integrated optical frequency combs with higher repetition rates could further increase the achievable communication rate by supporting higher symbol rates for intensity encoding.

    \begin{figure*}%[t]
	\centering
	\includegraphics[width= \linewidth]{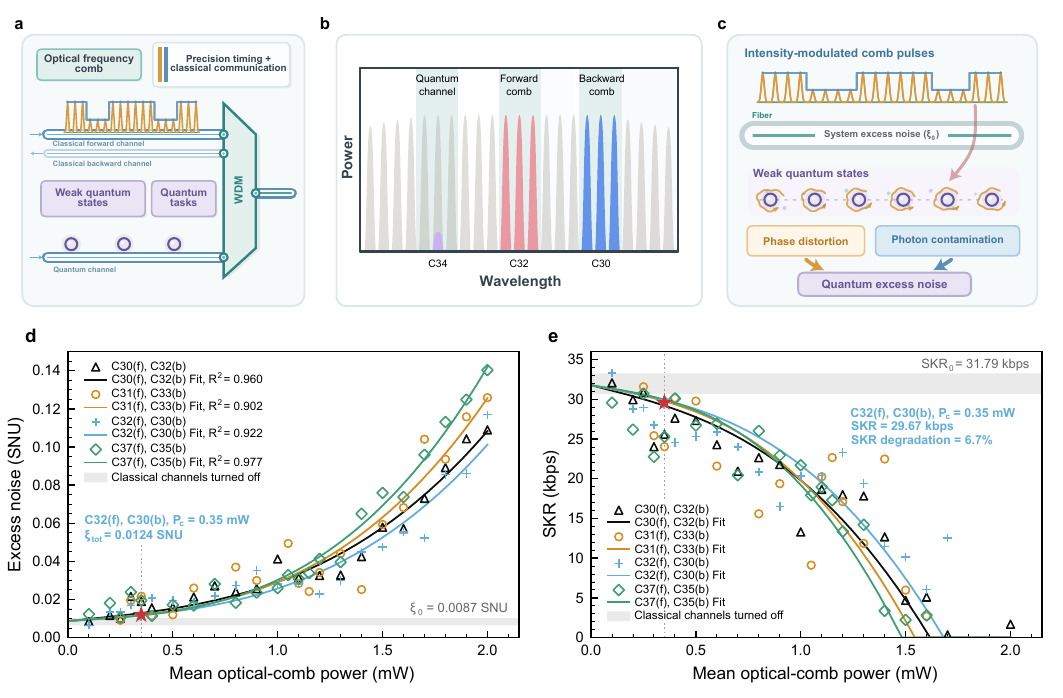}
	\caption{Integration of quantum tasks with the fused time-frequency communication infrastructure. a, Architecture for integrating quantum tasks. 
    b, Wavelength allocation of the classical and quantum channels. 
    c, Noise mechanisms. The broadband, high-peak-power optical frequency comb induces phase distortion through cross-phase modulation, while Raman scattering, four-wave mixing and residual optical leakage cause photon contamination. Together, these effects contribute to the excess noise of the quantum channel. 
    d, Excess noise versus the mean optical-comb power. 
    The labels f and b denote forward and backward comb propagation relative to the CV-QKD signal, respectively, with the quantum channel fixed at C34. 
    Black, yellow, blue and green curves correspond to C30(f), C32(b); C31(f), C33(b); C32(f), C30(b); and C37(f), C35(b), respectively. 
    $\xi_0$ is the mean system excess noise measured with the classical channels turned off, and the grey region indicates its measured fluctuation range. 
    e, SKR versus the mean optical-comb power, with the same colour coding as in d. 
    $\mathrm{SKR}_0$ and the grey region denote the mean and fluctuation range with the classical channels turned off, respectively. 
    }\label{Fig3}
    \end{figure*}

\subsection{Integration of quantum tasks through wavelength-division multiplexing}

The fusion of timing and classical communication infrastructures provides a unified optical platform for supporting future quantum networks, thereby mitigating resource constraints and improving scalability. However, practical quantum tasks rely on fragile quantum states, which generally require low-noise transmission conditions and cannot directly share a high-power classical carrier.

Unlike conventional continuous-wave (CW) classical channels, the fused carrier here is a broadband, high-peak-power pulse train composed of densely spaced, mutually coherent comb lines. These comb-specific temporal and spectral features make its coexistence with weak quantum states particularly demanding, which has remained largely unexplored.
Therefore, a key challenge is to integrate quantum task layers with the fused infrastructure while preserving the performance of quantum operations.
Here, we extend the optical-frequency-comb-enabled infrastructure to quantum tasks through wavelength-division multiplexing (WDM), as shown in Fig.~\ref{Fig3}a--b (see Methods for the CV-QKD experimental setup). In this architecture, the optical frequency comb serves as the carrier for the fused time-frequency communication infrastructure, while independent weak quantum states occupy dedicated wavelength channels and coexist within the same fibre link. Such a configuration preserves the spectral separation between classical and quantum signals while enabling them to share the same optical transmission infrastructure. 

The main limitation of this signal-level integration arises from noise transferred from classical optical channels into quantum channels, as illustrated in Fig.~\ref{Fig3}c.
The classical signals can degrade quantum performance through several mechanisms. Cross-phase modulation (XPM), enhanced by the high peak power of the comb, maps classical intensity fluctuations onto the quantum phase and causes phase distortion~\cite{du2020fwm,kumar2015coexistence}. Raman scattering and four-wave mixing (FWM) can generate photons within the quantum receiver band, and once in band, these photons cannot be removed by wavelength filtering and contribute to photon contamination~\cite{du2020fwm,kumar2015coexistence, ferreira2014raman}. Residual optical leakage due to finite channel isolation also introduces additional photon contamination~\cite{eriksson2019wdm,kumar2015coexistence,qi2010feasibility}.
For CV-QKD, these effects appear as excess noise and directly reduce the achievable secret key rate (SKR)~\cite{kumar2015coexistence,eriksson2019wdm,kiselev2021theoretical}. 
Moreover, compared with the conventional CW laser, this high-peak-power and broad-spectral-bandwidth light can further enhance nonlinear interactions, while its highly coherent spectral structure particularly facilitates FWM. In general, appreciable higher-order nonlinear contributions are indicated because of the complex spectral structure of the OFC carrier.

To quantify and optimize this coexistence, we investigate the dependence of quantum-channel performance on classical optical power, wavelength allocation and propagation direction. 
On this basis, we construct a power-dependent model for the excess noise, \begin{equation}
    \xi_{\mathrm{tot}}(P_\mathrm{c})\simeq \xi_0+c_1P_\mathrm{c}+c_2P_\mathrm{c}^2+c_3P_\mathrm{c}^3,
\end{equation}
where $P_\mathrm{c}$ is the mean optical-comb power, $\xi_0$ is the mean system excess noise measured with the classical channels turned off, %denotes the fitted system excess-noise term, 
and $c_1$, $c_2$ and $c_3$ are configuration-dependent coefficients that predominantly account for Raman scattering and optical leakage, XPM-induced phase distortion and FWM-induced photon contamination, respectively (see Supplementary Note~2 for the model formulation and derivation).  

This model quantitatively links the classical-channel conditions to the quantum excess noise and identifies optical power, wavelength allocation and propagation direction as the principal control parameters. 
As shown in Fig.~\ref{Fig3}d, the model is fitted to the measured excess-noise data for each channel configuration, and the excess noise increases at an accelerating rate with the mean optical-comb power for all channel allocations. 
Comparison of the fitted curves further reveals the effects of propagation direction and wavelength allocation. 
%方向选择
Specifically, two-way time--frequency transfer requires bidirectional comb transmission while the contributions of forward and backward Raman noise are different, resulting in the importance of selecting  propagation direction. At medium-to-long distances, as in our 100-km link, backward Raman scattering contributes more strongly than forward Raman scattering. 
By comparing the blue curve and black curve in Fig.~\ref{Fig3}d, we can see that the backward path benefits from a larger spectral separation from the quantum channel. 
%波长配置
Meanwhile, the comparison between the black curve and the yellow curve shows that increasing the spectral separation from the quantum channel reduces the overall noise contribution. 
At comparable spectral separations, placing the classical channels on the longer-wavelength side of quantum channel (C34) yields lower excess noise (yellow curve) than placing them on the shorter-wavelength side (green curve), consistent with weaker anti-Stokes than Stokes scattering into the quantum channel. 
Fig.~\ref{Fig3}e shows the corresponding SKR results, exhibiting the same dependence on wavelength allocation and propagation direction.
Hence, increasing the wavelength separation between classical and quantum channels reduces spectral overlap and suppresses noise transfer, while proper selection of propagation direction further minimizes Raman-induced noise. 
More importantly, there is always an appropriate power range where phase distortion and photon contamination are suppressed, keeping the excess noise low while maintaining the performance of the fused system. In general, by jointly optimizing propagation direction, wavelength allocation and optical power, the performance of the fused architecture can be optimized. 

We experimentally demonstrate the compatibility between the fused time-frequency communication infrastructure and a CV-QKD system over a 100-km fibre link, as indicated by the red stars in Fig.~\ref{Fig3}d--e. 
By optimizing the classical channel configuration, including optical power (0.35~mW), wavelength allocation and propagation direction (the blue curve: forward channel is C32 and backward channel is C30), the additional excess noise introduced into the quantum channel is suppressed to 0.0037 SNU, 
yielding an SKR of 29.67 kbps and thus resulting in a secret key rate degradation of 6.7$\%$ compared with the standalone quantum communication system (see Methods for the secret-key-rate calculation).

These results demonstrate that the optical-frequency-comb-enabled infrastructure can provide simultaneous support for precision timing, classical communication and quantum tasks, offering a practical pathway towards integrated and scalable quantum network deployment.

   \begin{figure*}[htp]%[t]
	\centering
	\includegraphics[width= \linewidth]{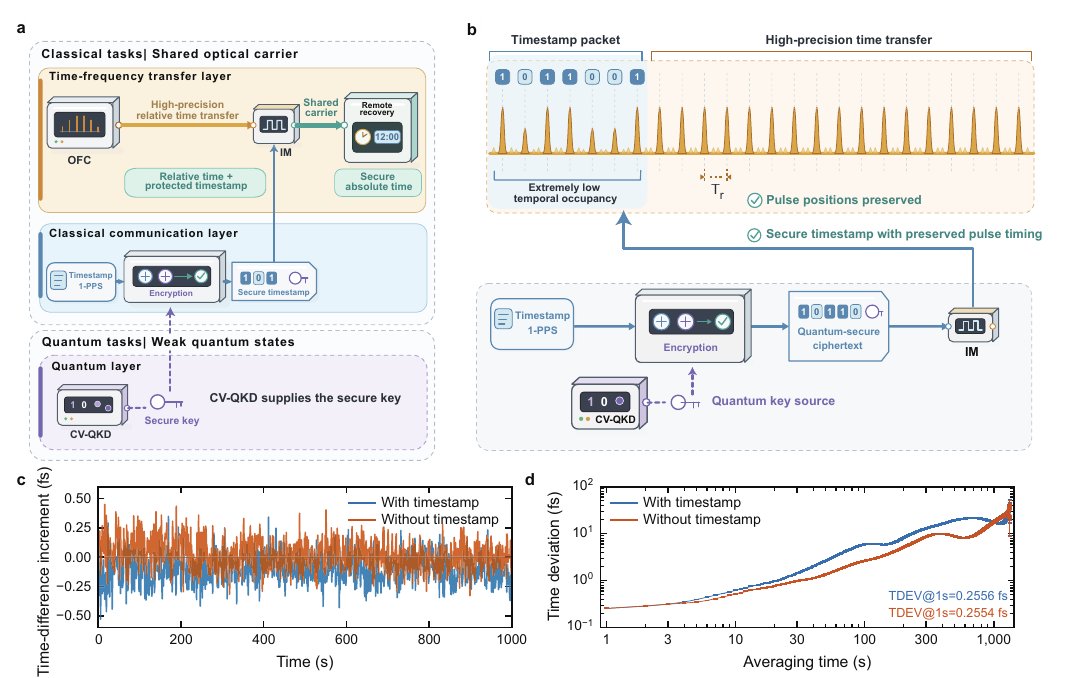}
	\caption{Secure absolute time synchronization enabled by the unified quantum network infrastructure. a, Unified quantum-network infrastructure for secure timestamp transmission. b, Secure integration mechanism for timestamp transmission. 
    c, Time-difference increments between adjacent acquisition points, showing the short-term timing fluctuations with and without timestamp transmission. 
    d,  Experimental results of timing stability. Error bars indicate one standard error of the TDEV estimates. The results obtained with and without timestamp transmission are identical at short time-scales. }\label{Fig4}
    \end{figure*}

\subsection*{Secure absolute time synchronization enabled by the unified quantum network infrastructure}

Beyond supporting quantum communication tasks, a practical quantum network requires reliable absolute synchronization among distributed nodes. 
The encoded sequences serving as timestamps carry absolute timing information, complementing the high-stability relative time transfer provided by the optical comb. 
Conventional absolute time synchronization relies on dedicated timing infrastructures and communication infrastructures, while security functions are typically implemented independently through cryptographic protocols to encrypt the timestamps. 
The separation between timing and security limits the ability to provide data--trustworthy synchronization in distributed quantum networks, where both temporal accuracy and information security are essential.

The unified infrastructure introduced here provides a new approach to secure timing services by combining precision time transfer with quantum-secured information exchange within the same optical network. 

To demonstrate this capability, we integrate the precision time transfer system with a CV-QKD channel through WDM.  As illustrated in Fig.~\ref{Fig4}a, in this architecture, the timing information is transferred using the optical-comb-based fused infrastructure, while the quantum communication channel provides the information-theoretically secure timestamps by applying one-time-pad encryption through a bitwise exclusive-OR operation using quantum-generated keys. 
As shown in Fig.~\ref{Fig4}b, the timestamp occupies only a negligible fraction of the transmitted data and therefore has virtually no effect on the stability of the time--frequency information. Hence, the coexistence of these functions enables absolute time synchronization with both high stability  and enhanced security with quantum-key protection, compared with conventional synchronization schemes that rely solely on classical encryption strategies. 

Experimentally, we demonstrate secure absolute time synchronization over a 100-km spooled fibre (see Methods for timestamp loading and timing-stability analysis). As shown in Fig.~\ref{Fig4}c, the amplitudes of time-difference increment remain nearly unchanged and are at the sub-femtosecond level, while maintaining secure timestamp exchange enabled by the quantum communication channel. The %stability of the 
fused system achieves a TDEV of 0.26 fs at 1~s, as shown in Fig.~\ref{Fig4}d. This performance remains consistent with that of the standalone precision timing system, indicating that the integration of security functions does not compromise timing stability.

   \begin{figure*}
	\centering
	\includegraphics[width= \linewidth]{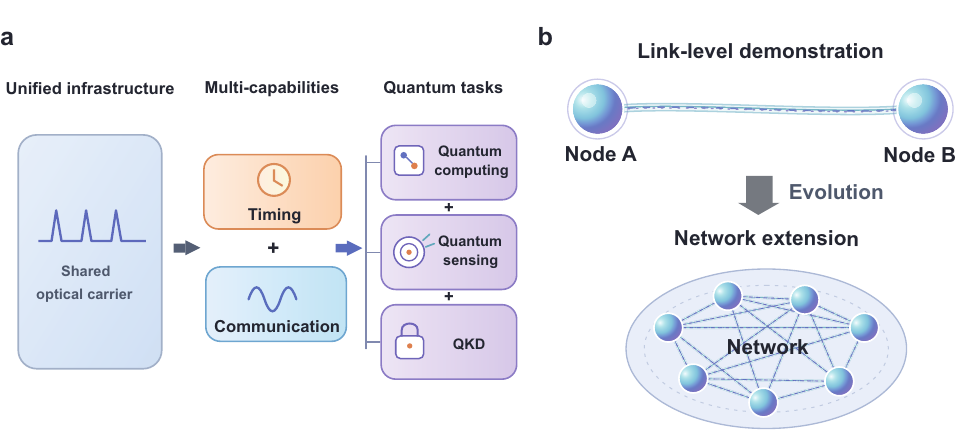}
	\caption{Towards scalable quantum networks with unified optical infrastructures. a, Prospective functionalities of the fused network. b, Extension from point-to-point links to network-level connectivity.}\label{Fig5}
    \end{figure*}

This demonstration highlights a new capability enabled by the unified infrastructure: precision timing, classical communication and quantum security can be jointly provided within a common optical network architecture. Such a capability is expected to play an important role in future distributed quantum computing, quantum sensing networks and large-scale quantum communication systems.

\section*{Discussion}

The realization of large-scale quantum networks requires the coordinated operation of multiple heterogeneous systems, including quantum communication channels, classical control networks and precision timing infrastructures~\cite{dixmatthews2021point}. Existing approaches typically rely on separately deployed systems, which increases hardware complexity and limits scalability. The unified infrastructure demonstrated here provides an alternative paradigm, where optical frequency combs enable the convergence of timing and communication functions while maintaining compatibility with weak quantum states.

A key advantage of this architecture is that it does not require replacing existing quantum technologies or communication infrastructures. Instead, it introduces an additional integration layer that enables quantum tasks to access shared timing and communication resources. Such an approach is compatible with existing fibre-optic networks and can potentially support diverse quantum applications, including distributed quantum computing, quantum sensing networks and large-scale quantum communication systems, as illustrated in Fig.~\ref{Fig5}a.

Several challenges remain for practical deployment. Extending the current demonstration to large-scale quantum networks shown in Fig.~\ref{Fig5}b will require improvements in integrated optical-frequency-comb sources, network-level synchronization protocols and long-distance quantum communication technologies. In particular, compact and robust comb generation, advanced channel management and improved suppression of nonlinear and scattering-induced noise will be important for metropolitan and intercity quantum networks. Future developments in integrated photonics may provide a pathway towards miniaturized and scalable implementations of the proposed architecture.

Beyond individual quantum links, the concept of infrastructure fusion may provide a general framework for future information networks, where precision timing, classical communication and quantum information processing are no longer treated as independent systems. 
As precise time–frequency dissemination also underpins integrated communication and navigation  networks, such fusion naturally extends to these systems, allowing them to be built on the same comb-based physical foundation. By establishing optical frequency combs as a common physical foundation for heterogeneous network functions, this work opens new opportunities for building secure, scalable and integrated quantum network infrastructures.

\section*{Methods}

\subsection*{Principle of dual-comb two-way time--frequency transfer and timing--communication fusion}

An optical frequency comb consists of a train of equally spaced femtosecond pulses in the time domain and a series of equally spaced comb lines in the frequency domain. At each site, the repetition frequencies of the combs are phase-locked to the local clock, while their carrier-envelope offset  frequencies are left free-running. The two combs are denoted as the clock comb with repetition frequency $f_r$ and the local oscillator (LO) comb with repetition frequency $f_r+\Delta f_r$, respectively. 
The clock comb maps the local timescale onto a stable optical pulse train, whose pulse-to-pulse spacing and pulse timing carry frequency and time information, respectively. Before transmission, classical communication data are encoded onto the intensity of the clock comb through intensity modulation while retaining its pulse timing. At the receiver, the same optical-comb signal is used for both data recovery and timing extraction, thereby fusing time--frequency transfer and classical communication on a shared optical carrier.

High-precision timing extraction is performed using dual-comb linear optical sampling (LOS). At each site, the LO comb interferes separately with the local clock comb and the received remote clock comb, generating the reference and target interferograms, respectively. 
The small repetition-rate offset between the two combs causes the relative pulse delay to scan from pulse to pulse, producing a time-stretched cross-correlation waveform. The single-step delay is
\begin{equation}
\delta\tau_{\mathrm{s}}
=
\left|
\frac{1}{f_r}
-
\frac{1}{f_r+\Delta f_r}
\right|
\simeq
\frac{\Delta f_r}{f_r^2},
\qquad
\Delta f_r\ll f_r .
\end{equation}
Accordingly, a physical time offset $\delta t$ is stretched on the laboratory timescale as
\begin{equation}
\delta t_{\mathrm{lab}}
=
\alpha\delta t,
\qquad
\alpha=\frac{f_r}{\Delta f_r},
\end{equation}
where $\alpha$ is the LOS time-stretching factor~\cite{yu2023time}. 
The LOS update period is
\begin{equation}
T_{\mathrm{LOS}}
=
\frac{1}{\Delta f_r}.
\end{equation}
This process maps the pulse cross-correlation onto the laboratory timescale, magnifying both its envelope width and the relative pulse delay by a factor of $\alpha$. The resulting LOS interferogram can be directly digitized by electronic acquisition systems for precise timing extraction.

For a reference or target interferogram $i$, the envelope $A_i(t)$ is first extracted using a Hilbert transform, and its arrival time is then determined from the envelope centroid~\cite{chen2024dualcomb},
\begin{equation}
t_i^{\mathrm{lab}}
=
\frac{
\displaystyle\int_{W} t A_i(t)\,\mathrm{d}t
}{
\displaystyle\int_{W} A_i(t)\,\mathrm{d}t
},
\qquad
i=1,2,3,...,
\end{equation}
where $W$ denotes the timing-extraction window. At sites A and B, the respective time differences on the laboratory timescale, $t_A^{\mathrm{lab}}$ and $t_B^{\mathrm{lab}}$, are obtained by comparing the centroid positions of the corresponding reference and target interferograms. The effective time differences are
\begin{equation}
\begin{aligned}
t_A
&=
\frac{
t_{A}^{\mathrm{lab}}}{\alpha},\qquad
t_B
&=
\frac{
t_{B}^{\mathrm{lab}}
}{\alpha}.
\end{aligned}
\end{equation}
Let $\Delta t_{AB}$ denote the clock offset at site B relative to site A, and let $\tau_{AB}$ and $\tau_{BA}$ denote the A-to-B and B-to-A propagation delays, respectively. The measured time intervals are then
\begin{equation}
t_A=\tau_{BA}-\Delta t_{AB},
\qquad
t_B=\tau_{AB}+\Delta t_{AB}.
\end{equation}
The two-way combination gives
\begin{equation}
\Delta t_{AB}
=
\frac{t_B-t_A}{2}
+
t_{\mathrm{NR}},
\qquad
t_{\mathrm{NR}}
=
\frac{\tau_{BA}-\tau_{AB}}{2},
\end{equation}
where $t_{\mathrm{NR}}$ is the non-reciprocal link delay. Under link reciprocity, that is, $\tau_{AB}\simeq\tau_{BA}$,
\begin{equation}
\Delta t_{AB}
\simeq
\frac{t_B-t_A}{2}.
\end{equation}
Consequently, the reciprocal propagation delay and its fluctuations are rejected as common-mode terms in the two-way measurement, leaving only the inter-site clock offset and residual link non-reciprocity. The fractional frequency offset is further obtained from
\begin{equation}
y_{AB}(t)
=
\frac{\mathrm{d}\Delta t_{AB}(t)}{\mathrm{d}t}.
\end{equation}
LOS resolves the relative clock offset within one pulse period at high resolution, while the integer-cycle ambiguity is resolved by the timestamp.

\subsection{Experimental architecture for timing--communication fusion}

The experimental architecture is shown in Fig.~\ref{Fig2}a. Sites A and B are connected by a 100-km spooled fibre. Comb A and Comb B serve as the clock combs at a repetition frequency of $100~\mathrm{MHz}$, whereas Comb $X_1$ and Comb $X_2$ serve as the corresponding LO combs at $100.001~\mathrm{MHz}$. At each site, a clock-referenced direct digital synthesizer (DDS) supplies the reference signals for phase-locking the comb repetition frequencies through proportional--integral--derivative (PID) feedback.

The clock-comb signals propagate in opposite directions over the shared link. At site A, an arbitrary waveform generator (AWG) drives an intensity modulator (IM) to encode classical data onto the transmitted output of Comb A. 
A local branch is split off before the IM to generate the reference interferogram. 
At site B, a beam splitter (BS) divides the received modulated signal between the LOS timing-extraction branch and the communication receiver. In the communication branch, the signal is amplified by an erbium-doped fibre amplifier (EDFA), photodetected and digitized by an analogue-to-digital converter (ADC) for data recovery.

At each site, the LO comb interferes separately with the local clock-comb output and the received remote clock-comb signal to generate the reference and target interferograms, respectively. These interferograms are digitized by ADCs, and the time intervals are extracted from their envelope centroids. The time intervals obtained at the two sites are then combined to determine the inter-site time offset using the procedure described in the preceding subsection.

To isolate the system-induced timing instability from the free-running offset between independent remote clocks, we employ a common-clock configuration in which the compared signals share the same frequency reference, so that the measured time difference reflects the residual time offset of the system.
The resulting time-difference sequence is then Kalman-filtered to suppress measurement noise.
The reported TDEV is used to characterize the stability of this residual time offset.

\subsection*{Experimental setup of CV-QKD}

We demonstrate the co-transmission of Gaussian-modulated coherent-state continuous-variable quantum key distribution (GMCS CV-QKD) signals and classical OFC signals over a single fibre link in the C band using dense WDM. Before multiplexing, the quantum signals are prepared with a target modulation variance $V_\mathrm{A}$, while the OFC signals are amplified by an erbium-doped fibre amplifier placed in the classical-comb path to compensate for the transmission loss. The quantum and amplified OFC signals are then multiplexed through the WDM system and transmitted over the shared fibre link.

At Alice, a narrow-linewidth CW laser operating at $1550.12~\mathrm{nm}$, corresponding to the C34 channel, serves as the optical carrier for quantum-state preparation. Coherent states are prepared using an FPGA-driven IQ modulator equipped with an integrated 16-bit digital-to-analog converter%(DAC)
, generating GMCS at a repetition rate of $100~\mathrm{MHz}$. A pilot-sequential scheme is employed to alternately transmit reference pulses and quantum signal pulses, maintaining the quantum signals at the weak-light level while providing the reference required for phase recovery. Finally, a variable optical attenuator is employed to fine-tune the modulation variance of the transmitted GMCS to the optimal value $V_\mathrm{A}$.

At Bob, quantum-state measurements are performed via heterodyne detection using an independent local oscillator. The detector output is digitized in real time by an analog-to-digital converter operating at $1~\mathrm{GSa/s}$. The acquired samples are subsequently subjected to real-time digital signal processing for phase recovery and compensation to recover the quantum signals. The recovered data are then subjected to parameter estimation, information reconciliation, and privacy amplification to extract the final secret key.

\subsection*{Secret key rate calculation of CV-QKD}

For a GMCS CV-QKD system with reverse reconciliation, the secret key rate in the asymptotic regime  can be expressed as~\cite{pirandola2020advances,laudenbach2018cvqkd}: 
\begin{equation}
K\left[\xi_\mathrm{tot}\left(P_\mathrm{c}\right),T\right]
=
R_\mathrm{eff}
\left\{
\beta I_\mathrm{AB}
-
\chi_\mathrm{BE}
\right\}_{+},
\label{eq:cvqkd_skr}
\end{equation}
where $R_\mathrm{eff}$ denotes the effective quantum symbol rate, $\beta$ is the reconciliation efficiency, $I_\mathrm{AB}$ represents the Shannon mutual information between Alice and Bob, and $\chi_\mathrm{BE}$ denotes the maximum information accessible to Eve.

In this experiment, the classical optical signals and CV-QKD signals are co-transmitted over the same fibre link. The classical optical signals affect the CV-QKD system primarily by introducing additional channel noise. Denoting the average classical optical power by $P_\mathrm{c}$, the experimentally measured total excess noise, denoted by $\xi_\mathrm{tot}\left(P_\mathrm{c}\right)$, is substituted into Eq.~\eqref{eq:cvqkd_skr}. Accordingly, the total channel noise referred to the channel input can be expressed as
\[
\chi_\mathrm{line}\left(P_\mathrm{c}\right)
=
\frac{1}{T}
-
1
+
\xi_\mathrm{tot}\left(P_\mathrm{c}\right).
\]

For heterodyne detection, the detection-added noise is given by
\[
\chi_\mathrm{het}
=
\frac{1+(1-\eta)+2v_\mathrm{el}}{\eta},
\]
where $\eta$ denotes the detection efficiency and $v_\mathrm{el}$ is the normalized electronic noise of the detector. The total noise referred to the channel input can therefore be written as
\[
\chi_\mathrm{tot}\left(P_\mathrm{c}\right)
=
\chi_\mathrm{line}\left(P_\mathrm{c}\right)
+
\frac{\chi_\mathrm{het}}{T}.
\]

Considering the dependence of the total noise on the average classical optical power $P_\mathrm{c}$, the mutual information between Alice and Bob is expressed as
\[
I_\mathrm{AB}\left(P_\mathrm{c}\right)
=
\log_{2}
\left[
\frac{
V+\chi_\mathrm{tot}\left(P_\mathrm{c}\right)
}{
1+\chi_\mathrm{tot}\left(P_\mathrm{c}\right)
}
\right],
\]
where $V=V_\mathrm{A}+1$, and $V_\mathrm{A}$ denotes the modulation variance.

The average classical optical power $P_\mathrm{c}$, effective quantum symbol rate $R_\mathrm{eff}$, reconciliation efficiency $\beta$, modulation variance $V_\mathrm{A}$, channel transmittance $T$, electronic noise $v_\mathrm{el}$, and detection efficiency $\eta$ are experimentally measured or predetermined system parameters. The Holevo information between Bob and Eve is given by
\[
\begin{aligned}
\chi_\mathrm{BE}\left(P_\mathrm{c}\right)
={}&
G\left[
\frac{\lambda_{1}\left(P_\mathrm{c}\right)-1}{2}
\right]
+
G\left[
\frac{\lambda_{2}\left(P_\mathrm{c}\right)-1}{2}
\right] \\
&-
G\left[
\frac{\lambda_{3}\left(P_\mathrm{c}\right)-1}{2}
\right]
-
G\left[
\frac{\lambda_{4}\left(P_\mathrm{c}\right)-1}{2}
\right]\\
&-
G\left[
\frac{\lambda_{5}\left(P_\mathrm{c}\right)-1}{2}
\right],
\end{aligned}
\]
where $G(x)$ is defined as
\[
G(x)
=
(x+1)\log_{2}(x+1)
-
x\log_{2}x.
\]

We further define
\[
A\left(P_\mathrm{c}\right)
=
V^{2}(1-2T)
+
2T
+
T^{2}
\left[
V+\chi_\mathrm{line}\left(P_\mathrm{c}\right)
\right]^{2},
\]

\[
B\left(P_\mathrm{c}\right)
=
T^{2}
\left[
V\chi_\mathrm{line}\left(P_\mathrm{c}\right)+1
\right]^{2},
\]

\[
\begin{aligned}
C\left(P_\mathrm{c}\right)
=&
\frac{1}{
T^{2}
\left[
V+\chi_\mathrm{tot}\left(P_\mathrm{c}\right)
\right]^{2}
}
\Bigg\{
A\left(P_\mathrm{c}\right)\chi_\mathrm{het}^{2}
+
B\left(P_\mathrm{c}\right)
+
1 \\
&+
2\chi_\mathrm{het}
\left[
V\sqrt{B\left(P_\mathrm{c}\right)}
+
T\left(
V+\chi_\mathrm{line}\left(P_\mathrm{c}\right)
\right)
\right] \\
&+
2T\left(V^{2}-1\right)
\Bigg\},
\end{aligned}
\]

\[
D\left(P_\mathrm{c}\right)
=
\left[
\frac{
V+\sqrt{B\left(P_\mathrm{c}\right)}\chi_\mathrm{het}
}{
T\left[
V+\chi_\mathrm{tot}\left(P_\mathrm{c}\right)
\right]
}
\right]^{2}.
\]

Here, the symplectic eigenvalues can then be expressed as functions of $P_\mathrm{c}$:
\[
\lambda_{1}\left(P_\mathrm{c}\right)
=
\sqrt{
\frac{
A\left(P_\mathrm{c}\right)
+
\sqrt{
A^{2}\left(P_\mathrm{c}\right)
-
4B\left(P_\mathrm{c}\right)
}
}{2}
},
\]

\[
\lambda_{2}\left(P_\mathrm{c}\right)
=
\sqrt{
\frac{
A\left(P_\mathrm{c}\right)
-
\sqrt{
A^{2}\left(P_\mathrm{c}\right)
-
4B\left(P_\mathrm{c}\right)
}
}{2}
},
\]

\[
\lambda_{3}\left(P_\mathrm{c}\right)
=
\sqrt{
\frac{
C\left(P_\mathrm{c}\right)
+
\sqrt{
C^{2}\left(P_\mathrm{c}\right)
-
4D\left(P_\mathrm{c}\right)
}
}{2}
},
\]

\[
\lambda_{4}\left(P_\mathrm{c}\right)
=
\sqrt{
\frac{
C\left(P_\mathrm{c}\right)
-
\sqrt{
C^{2}\left(P_\mathrm{c}\right)
-
4D\left(P_\mathrm{c}\right)
}
}{2}
},
\]

\[
\lambda_{5}\left(P_\mathrm{c}\right)
=
1.
\]

By substituting the above expressions into the secret key rate formula, the dependence of the secret key rate $K$ on the average classical optical power $P_\mathrm{c}$ can be obtained.

\subsection*{Timestamp loading and timing-stability analysis}

In the secure absolute time-synchronization experiment, a 1-PPS signal triggers one 76-bit timestamp packet per second (see Supplementary Note~3 for the timestamp format). The timestamp is protected by quantum keys and loaded onto the optical-comb carrier through intensity modulation at the bit rate of $100~\mathrm{Mbit\,s^{-1}}$, using NRZ coding with $D = 50\%$ and $m = 1$. 
Each timestamp has a duration of $760~\mathrm{ns}$, corresponding to an average timestamp transmission rate of $76~\mathrm{bit\,s^{-1}}$ and a temporal occupancy of only $7.6\times10^{-7}$. This extremely low temporal occupancy makes the impact of timestamp loading on timing stability negligible.

For Fig.~\ref{Fig4}c, the time-difference increments are obtained by differencing adjacent samples and subtracting the global mean of the full increment sequence. The nearly identical fluctuation amplitudes with and without timestamp loading further confirm that timestamp modulation has negligible impact on time--frequency-transfer stability.

\section*{Data availability}
The data that support the findings of
this study are not publicly available. The data are available
from the authors upon reasonable request.

\section*{Code availability}
The codes that support the findings of this study are available on reasonable request from the corresponding authors.

\section*{Acknowledgements}

This work was supported by the National Natural Science Foundation of China (62201012, 62571006) and the Fundamental Research Funds for the Central Universities (2026CX01008).

\section*{Author contributions}

Y.L., K.X., Z.C. and H.G. conceived the idea and initiated the project. Y.L., X.W., Z.C. developed the concept of the optical-frequency-comb-enabled unified infrastructure, and Y.L., Z.R., Z.C., K.X., H.G. designed the overall experimental architecture. Y.L., Z.R., Y.F.Z., Z.L., X.W., Z.C. designed and constructed the time--frequency transfer system, and Y.L., Z.R., Y.K.Z., Z.L., Y.F.Z., Y.H.S., and Z.C.  performed the experiments. Y.L. and Z.R. developed the theoretical model of modulation-induced timing noise, carried out the numerical simulations and analysed the timing-stability data. Y.L., Y.S., L.Y., and Z.C. designed the classical communication scheme and performed the communication experiments. J.M., Y.H.S., and X.W. designed, constructed and operated the CV-QKD system. Y.L., J.M. and Z.L. performed the CV-QKD-related experiments in the fused system. Y.L. and Z.C. wrote the original manuscript. All authors participated in data analysis and interpretation, reviewed and edited the manuscript, and approved its final version.

\section*{Competing interests}

The authors declare no competing interests.

% 补充材料
\clearpage
\onecolumngrid
\section*{Supplementary Information}

\setcounter{equation}{0}
\renewcommand{\theequation}{S\arabic{equation}}
\renewcommand{\theHequation}{supp.\arabic{equation}}

\subsection*{Supplementary Note 1. Analysis of the effect of intensity modulation on timing stability}

To analyze the dependence of linear optical sampling (LOS) timing stability%on modulation frequency and modulation depth
, we consider a single-frequency sinusoidal intensity modulation and evaluate its response under centroid-based timing extraction. Practical modulation waveforms can be decomposed into sinusoidal components, making single-frequency analysis a natural basis for evaluating their impact on timing stability and be able to serve as a proof of principle. Also, we used periodic sequences for principle validation in experiments to  ensure a fair comparison among different communication strategies.
%Also, the sequence we used for principle validation in experiments is periodic to 
%follow the controlled variable method and 
%ensure a fair comparison among different communication strategies. %This formulation isolates the two parameters of interest and allows the modulation-induced centroid shift within the finite timing-extraction window to be obtained analytically. 
%The resulting timing-error spectrum is then mapped to TDEV, providing the theoretical basis for Fig.~2f.
Let $t$ denote the time on the laboratory-time axis, and let the signal intensity be
\[
P(t)=P_0\left[1+m\sin(2\pi f_\mathrm{m} t)\right],\qquad 0\le m<1.
\]

Let the center of the timing-extraction window be $t_\mathrm{c}$. For the symmetric unmodulated reference, the centroid is
\[
C_0(t_\mathrm{c})=t_\mathrm{c}.
\]
%To analyze the effect of intensity modulation on LOS timing stability, 
%the problem can be reduced to the stability analysis of the centroid position of a sinusoidal signal under centroid-based timing extraction, with the analysis restricted to the window selected for LOS centroid estimation. 
%Let $t$ denote the time on the laboratory-time axis, and let the sinusoidal signal intensity be
%\[
%P(t)=P_{0}\left[1+m\sin(2\pi f_\mathrm{m}t)\right],
%\qquad 0\leq m<1.
%\]
%Let the center of the analysis window be $t_\mathrm{c}$. For a %symmetric waveform, the ideal centroid is
%\[
%C_{0}(t_\mathrm{c})=t_\mathrm{c}.
%\]
The centroid position of the signal is
\[
C(t_\mathrm{c})
=
\frac{
\displaystyle\int_{t_\mathrm{c}-W/2}^{t_\mathrm{c}+W/2}tP(t)\,\mathrm{d}t
}{
\displaystyle\int_{t_\mathrm{c}-W/2}^{t_\mathrm{c}+W/2}P(t)\,\mathrm{d}t
}.
\]
The relative time on the oscilloscope is defined as
\[
u=t-t_\mathrm{c},
\qquad
-\frac{W}{2}\leq u\leq\frac{W}{2},
\]
and the modulation-induced time offset is
\[
x(t_\mathrm{c})=C(t_\mathrm{c})-t_\mathrm{c}.
\]

Rewriting the denominator in terms of $u$ and expanding gives
\begin{align*}
E(t_\mathrm{c})
&=
\int_{-W/2}^{W/2}P(t_\mathrm{c}+u)\,\mathrm{d}u
=
P_{0}\int_{-W/2}^{W/2}
\Big[
1+m\sin(2\pi f_\mathrm{m}t_\mathrm{c})\cos(2\pi f_\mathrm{m}u)
+m\cos(2\pi f_\mathrm{m}t_\mathrm{c})\sin(2\pi f_\mathrm{m}u)
\Big]\mathrm{d}u
\\&=
P_{0}W
\left[
1+
m\sin(2\pi f_\mathrm{m}t_\mathrm{c})
\frac{\sin(\pi f_\mathrm{m}W)}{\pi f_\mathrm{m}W}
\right].
\end{align*}

The numerator is
\begin{align*}
M(t_\mathrm{c})
&=
\int_{-W/2}^{W/2}
(t_\mathrm{c}+u)P(t_\mathrm{c}+u)\,\mathrm{d}u
=
t_\mathrm{c}
\int_{-W/2}^{W/2}
P(t_\mathrm{c}+u)\,\mathrm{d}u
+
\int_{-W/2}^{W/2}
uP(t_\mathrm{c}+u)\,\mathrm{d}u
=
t_\mathrm{c}E(t_\mathrm{c})
+
\int_{-W/2}^{W/2}
uP(t_\mathrm{c}+u)\,\mathrm{d}u.
\end{align*}

The centroid shift is therefore
\begin{equation}
x(t_\mathrm{c})
=
C(t_\mathrm{c})-t_\mathrm{c}
=
\frac{
\displaystyle\int_{-W/2}^{W/2}
uP(t_\mathrm{c}+u)\,\mathrm{d}u
}{
\displaystyle\int_{-W/2}^{W/2}
P(t_\mathrm{c}+u)\,\mathrm{d}u
}.
\label{eq:centroid_shift}
\end{equation}

Substituting $P(t_\mathrm{c}+u)$ gives
\begin{align*}
\int_{-W/2}^{W/2}
uP(t_\mathrm{c}+u)\,\mathrm{d}u
&=
P_{0}
\int_{-W/2}^{W/2}
u
\Big[
1
+m\sin(2\pi f_\mathrm{m}t_\mathrm{c})\cos(2\pi f_\mathrm{m}u)
+m\cos(2\pi f_\mathrm{m}t_\mathrm{c})\sin(2\pi f_\mathrm{m}u)
\Big]\mathrm{d}u
\\
&=0+
P_{0}m\cos(2\pi f_\mathrm{m}t_\mathrm{c})
\int_{-W/2}^{W/2}
u\sin(2\pi f_\mathrm{m}u)\,\mathrm{d}u
\\
&=
-P_{0}W\frac{m}{2\pi f_\mathrm{m}}
\left[
\cos(\pi f_\mathrm{m}W)
-
\frac{\sin(\pi f_\mathrm{m}W)}{\pi f_\mathrm{m}W}
\right]
\cos(2\pi f_\mathrm{m}t_\mathrm{c}).
\end{align*}

The exact centroid offset is therefore
\begin{equation}
\begin{aligned}
x_{\mathrm{strict}}(t_\mathrm{c})
&=
-\frac{m}{2\pi f_\mathrm{m}}
\left[
\cos(\pi f_\mathrm{m}W)
-
\frac{\sin(\pi f_\mathrm{m}W)}{\pi f_\mathrm{m}W}
\right]
\cdot
\frac{
\cos(2\pi f_\mathrm{m}t_\mathrm{c})
}{
1+
m\dfrac{\sin(\pi f_\mathrm{m}W)}{\pi f_\mathrm{m}W}
\sin(2\pi f_\mathrm{m}t_\mathrm{c})
}.
\end{aligned}
\label{eq:strict_centroid_shift}
\end{equation}

Let
$
B
=
-\frac{m}{2\pi f_\mathrm{m}}
\left[
\cos(\pi f_\mathrm{m}W)
-
\frac{\sin(\pi f_\mathrm{m}W)}{\pi f_\mathrm{m}W}
\right],
q
=
m\frac{\sin(\pi f_\mathrm{m}W)}{\pi f_\mathrm{m}W},
\psi
=2\pi f_\mathrm{m}t_\mathrm{c}.$
Equation~\eqref{eq:strict_centroid_shift} then becomes
$
x_{\mathrm{strict}}(\psi)
=
B\frac{\cos\psi}{1+q\sin\psi}.
$
Further defining
$\theta=\psi-\frac{\pi}{2}$
gives
\[
x_{\mathrm{strict}}(\theta)
=
-B\frac{\sin\theta}{1+q\cos\theta}.
\]

Since $|q|<1$, we define
$
\rho=
\frac{q}{1+\sqrt{1-q^{2}}},
$
such that
$
|\rho|<1,
q=\frac{2\rho}{1+\rho^{2}}.
$
Using the identity
\[
\sum_{k=1}^{\infty}
r^{k}\sin(k\theta)
=
\frac{r\sin\theta}
{1-2r\cos\theta+r^{2}},
\qquad |r|<1,
\]
and taking $r=-\rho$, we obtain
$
\sum_{k=1}^{\infty}
(-\rho)^{k}\sin(k\theta)
=
-\frac{\rho\sin\theta}
{1+2\rho\cos\theta+\rho^{2}}
=
-\frac{\rho\sin\theta}
{\dfrac{2\rho}{q}
\left(1+q\cos\theta\right)}.
$
Hence,
\[
-\frac{\sin\theta}{1+q\cos\theta}
=
\frac{2}{q}
\sum_{k=1}^{\infty}
(-\rho)^{k}\sin(k\theta).
\]
Therefore,
\[
x_{\mathrm{strict}}(\theta)
=
\frac{2B}{q}
\sum_{k=1}^{\infty}
(-\rho)^{k}\sin(k\theta),
\quad q\neq0,
\]
or equivalently,
\[
x_{\mathrm{strict}}(t_\mathrm{c})
=
\sum_{k=1}^{\infty}
C_{k}(f_\mathrm{m},m,W)
\sin\left[
k\left(
2\pi f_\mathrm{m}t_\mathrm{c}-\frac{\pi}{2}
\right)
\right],
\]
where
$
C_{k}(f_\mathrm{m},m,W)
=
\frac{2B}{q}(-\rho)^{k}.
$

Defining the amplitude of the $k$th harmonic as
$
A_{k}=|C_{k}|,
$
we have
\[
x_{\mathrm{strict}}(t_\mathrm{c})
=
\sum_{k=1}^{\infty}
A_{k}(f_\mathrm{m},m,W)
\sin\left(
2\pi kf_\mathrm{m}t_\mathrm{c}+\phi_{k}
\right),
\]
where
\begin{equation}
\begin{aligned}
A_{k}(f_\mathrm{m},m,W)
&=
\frac{
\left|
\cos X-\operatorname{sinc}(X)
\right|
}{
\pi f_\mathrm{m}\left|\operatorname{sinc}(X)\right|
}
\left[
\frac{
\left|m\operatorname{sinc}(X)\right|
}{
1+\sqrt{
1-m^{2}\operatorname{sinc}^{2}(X)
}
}
\right]^{k},
\end{aligned}
\label{eq:harmonic_amplitude}
\end{equation}
with
$
X=\pi f_\mathrm{m}W,
\operatorname{sinc}(X)=\frac{\sin X}{X}.
$
The initial phase of the $k$th harmonic is
$
\phi_{k}
=
-\frac{k\pi}{2}
+
\frac{1-\operatorname{sgn}(C_{k})}{2}\pi
\pmod{2\pi}.
$
For $q=0$, the corresponding expressions are understood in the continuous limit $q\rightarrow0$.

For a general timing-error process $x(t)$, the TDEV is obtained from its one-sided power spectral density $S_{x}(f)$ and the TDEV transfer function $H_{\mathrm{TDEV}}(f;\tau)$ as~\cite{riley2008handbook}
\begin{equation}
\sigma_{x}^{2}(\tau)
=
\int_{0}^{\infty}
S_{x}(f)
\left|
H_{\mathrm{TDEV}}(f;\tau)
\right|^{2}
\mathrm{d}f,
\label{eq:tdev_psd}
\end{equation}
where
$
\left|
H_{\mathrm{TDEV}}(f;\tau)
\right|
=
\sqrt{\frac{8}{3}}
\left|
\frac{
\sin^{3}(\pi f\tau)
}{
\pi f\tau
}
\right|,
$
and
$
S_{x}(f)
=
\sum_{k=1}^{\infty}
\frac{
A_{k}^{2}(f_\mathrm{m},m,W)
}{2}
\delta(f-kf_\mathrm{m}).
$

Substitution into Eq.~\eqref{eq:tdev_psd} gives
\begin{align*}
\sigma_{x}^{2}(\tau)
&=
\int_{0}^{\infty}
\sum_{k=1}^{\infty}
\frac{
A_{k}^{2}(f_\mathrm{m},m,W)
}{2}
\delta(f-kf_\mathrm{m})
\cdot
\left|
H_{\mathrm{TDEV}}(f;\tau)
\right|^{2}
\mathrm{d}f.
\end{align*}

Using the Dirac-delta identity
$
\int_{0}^{\infty}
\delta(f-kf_\mathrm{m})
\left|
H_{\mathrm{TDEV}}(f;\tau)
\right|^{2}
\mathrm{d}f
=
\left|
H_{\mathrm{TDEV}}(kf_\mathrm{m};\tau)
\right|^{2},
$
the integral reduces to the discrete sum
$
\sigma_{x}^{2}(\tau;f_\mathrm{m},m,W)
=
\sum_{k=1}^{\infty}
\frac{
A_{k}^{2}(f_\mathrm{m},m,W)
}{2}
\left|
H_{\mathrm{TDEV}}(kf_\mathrm{m};\tau)
\right|^{2},
$
that is,
\begin{align*}
\sigma_{x}^{2}(\tau;f_\mathrm{m},m,W)
&=
\frac{4}{3}
\sum_{k=1}^{\infty}
A_{k}^{2}(f_\mathrm{m},m,W)
\left[
\frac{
\left|\sin^{3}(\pi kf_\mathrm{m}\tau)\right|
}{
\pi kf_\mathrm{m}\tau
}
\right]^{2}.
\end{align*}

The TDEV is therefore
\begin{equation}
\begin{aligned}
\sigma_{x}(\tau;f_\mathrm{m},m,W)
&=
\Bigg\{
\frac{4}{3}
\sum_{k=1}^{\infty}
A_{k}^{2}(f_\mathrm{m},m,W)
\left[
\frac{
\left|\sin^{3}(\pi kf_\mathrm{m}\tau)\right|
}{
\pi kf_\mathrm{m}\tau
}
\right]^{2}
\Bigg\}^{1/2}.
\end{aligned}
\label{eq:tdev_final}
\end{equation}

\subsection*{Supplementary Note 2. Excess-noise model for CV-QKD}

After fusion of the communication and time-frequency layers, the impact of the classical channels on the quantum channel can be modeled as
\begin{equation}
\xi_{\mathrm{tot}}(P_\mathrm{c})
\simeq
\xi_{0}
+c_{1}P_\mathrm{c}
+c_{2}P_\mathrm{c}^{2}
+c_{3}P_\mathrm{c}^{3},
\label{eq:excess_noise_model}
\end{equation}
where the dominant contributions to $c_{1}$ are $\xi_{\mathrm{leak}}$ and $\xi_{\mathrm{Ram}}$, while $c_{2}$ and $c_{3}$ predominantly arise from $\xi_{\mathrm{XPM}}$ and $\xi_{\mathrm{FWM}}$, respectively. Excess noise is introduced through two mechanisms. First, it can be regarded as arising from additional photons detected by the receiver. Since
\[
\xi \propto \langle N\rangle\propto\frac{P}{h\nu}
\]
within a given detection window, the problem reduces to determining the dependence of the corresponding noise power on $P_\mathrm{c}$. Second, classical light induces changes in the fiber refractive index and thereby degrades the coherence of the quantum light, which is mainly manifested through cross-phase modulation (XPM).

The system excess-noise term $\xi_0$ can be measured directly. For the residual optical-leakage noise $\xi_{\mathrm{leak}}$, 
\[
P_{\mathrm{leak}}=
\sum_{i}10^{-I_{i}/10}P_{i},
\qquad
\xi_{\mathrm{leak}}
=
\kappa_{\mathrm{rx}}P_{\mathrm{leak}},
\]
where $I_{i}$ is the isolation between the $i$th classical channel and the quantum channel, and $\kappa_{\mathrm{rx}}$ is the conversion factor from optical power to shot-noise units (SNU). Therefore,
\[
\xi_{\mathrm{leak}}
\propto
P_{\mathrm{leak}}
\propto
P_\mathrm{c}.
\]

For $\xi_{\mathrm{Ram}}$, spontaneous Raman scattering is the dominant noise source, whereas stimulated Raman scattering is not dominant. Summarizing the spontaneous Raman scattering noise models in the literature~\cite{ferreira2014raman,kumar2015coexistence,eriksson2019wdm,iqbal2025sdn}, we obtain
\begin{equation}
\begin{aligned}
\xi_{\mathrm{Ram}}
&=
K_{\mathrm{Ram}}
\sum_{i}
\beta_{\sigma_{i}}(f_{q}-f_{i};f_{q})
\left[
P_{i,\mathrm{fwd}}Le^{-\alpha L}
+
P_{i,\mathrm{bwd}}
\frac{1-e^{-2\alpha L}}{2\alpha}
\right],
\\
\sigma_{i}
&=
\begin{cases}
\mathrm{AS}, & f_{q}>f_{i},\\
\mathrm{S},  & f_{q}<f_{i}.
\end{cases}
\end{aligned}
\label{eq:raman_noise}
\end{equation}
Here, $K_{\mathrm{Ram}}$ absorbs the effective linear coefficient and unit conversion and depends on wavelength and transmittance. $\beta_{\sigma_{i}}=\beta_{\sigma_{i}}(f_{q}-f_{i};f_{q})$ is the spontaneous Raman scattering coefficient and depends on the channel position. $\sigma_{i}=\mathrm{S}$ or $\mathrm{AS}$ indicates whether the quantum channel lies on the Stokes or anti-Stokes side of the corresponding classical channel. In general, $\beta_{\mathrm{S}}$ is larger, indicating that Stokes scattering is more probable than anti-Stokes scattering. $L$ is the transmission distance. The first term in the brackets represents forward Raman-scattering noise, which first increases and then decreases with transmission distance, whereas the second term represents backward Raman-scattering noise, which increases monotonically and gradually saturates with distance. This relation can therefore be used to select the wavelength-channel positions and propagation directions.

For Kerr-nonlinearity-induced noise, the dominant effects are cross-phase modulation (XPM), self-phase modulation (SPM), and four-wave mixing (FWM). Because the quantum signal is extremely weak, SPM can be neglected. XPM induces a phase shift in the quantum channel~\cite{kumar2015coexistence},
\begin{equation}
\phi_{\mathrm{XPM}}
=
\frac{4\pi n_{2}L_{\mathrm{eff}}}
{\lambda_{c}A_{\mathrm{eff}}}
\left(
P_\mathrm{c}^{\mathrm{s}}
-
P_\mathrm{c}^{\mathrm{r}}
\right),
\label{eq:xpm_phase}
\end{equation}
where $n_{2}$ is the nonlinear refractive index, $L_{\mathrm{eff}}$ is the effective length, $\lambda_{c}$ is the classical-channel wavelength, and $A_{\mathrm{eff}}$ is the effective mode area. $P_\mathrm{c}^{\mathrm{s}}$ and $P_c^{\mathrm{r}}$ denote the classical optical powers temporally overlapping with the quantum-signal and transmitted reference pulses, respectively. When the random phase perturbation $\delta\phi_{\mathrm{XPM}}$ can be treated as a small quantity,
\[
\xi_{\mathrm{XPM}}
\propto
V_{A}\operatorname{Var}\!\left(\phi_{\mathrm{XPM}}\right).
\]
Considering
$
P_\mathrm{c}^{\mathrm{s}}
-
P_\mathrm{c}^{\mathrm{r}}
\propto P_\mathrm{c},
$
we have
$
\phi_{\mathrm{XPM}}\propto P_\mathrm{c},
$
and therefore
\[
\xi_{\mathrm{XPM}}
\propto
\operatorname{Var}\!\left(\phi_{\mathrm{XPM}}\right)
\propto
\operatorname{Var}(P_\mathrm{c})
\propto
P_\mathrm{c}^{2}.
\]

FWM increases the noise by introducing additional photons~\cite{du2020fwm}. Under the slowly varying envelope approximation, the FWM-generated field amplitude is
\[
A_{ijk\rightarrow q}^{\mathrm{FWM}}
\propto
\gamma A_{i}A_{j}A_{k}^{*}
L_{\mathrm{eff}}
\Phi(\Delta\beta).
\]
Its squared magnitude gives the corresponding power,
\[
P_{ijk\rightarrow q}^{\mathrm{FWM}}
\propto
\gamma^{2}
P_{i}P_{j}P_{k}
\left|\Phi(\Delta\beta)\right|^{2},
\]
and the corresponding additive noise is
$
\xi_{\mathrm{FWM}}
=
\sum_{i,j,k}
C_{ijk}P_{i}P_{j}P_{k}.
$
If the powers in different spectral bands vary synchronously with the classical optical power, then
\[
\xi_{\mathrm{FWM}}
\simeq
c_{3}P_\mathrm{c}^{3}\propto P_\mathrm{c}^{3}.
\]

\subsection*{Supplementary Note 3. Timestamp packet format and loading}

For the secure absolute time-synchronization experiment, each timestamp packet consists of a fixed 12-bit preamble followed by a 64-bit protected timestamp payload,
\begin{equation}
\mathbf{b}_{\mathrm{ts}}
=
\left[
\mathbf{b}_{\mathrm{pre}}^{(12)}
\|
\mathbf{b}_{\mathrm{pay}}^{(64)}
\right],
\qquad
\mathbf{b}_{\mathrm{pre}}^{(12)}
=
010101010101.
\end{equation}
The alternating preamble provides a distinctive pattern for identifying the packet boundary and aligning the timestamp frame. The 64-bit payload carries the protected timestamp information used to associate the high-precision relative timing with its corresponding absolute-time label.

\bibliographystyle{apsrev4-2}
\bibliography{nsr_sample}
\end{document}